\documentclass[aps,pra,twocolumn]{revtex4-1}

\usepackage{graphicx}
\usepackage{amsmath}
\usepackage{amsfonts}
\usepackage{staves}
\usepackage{textcomp}
\usepackage{MnSymbol}
\usepackage{mathrsfs}
\usepackage{multirow}
\usepackage{fancyhdr}
\usepackage{bbm}
\usepackage{dsfont}
\usepackage{upgreek}

\begin{document}

\title{Bichromatic Driving of a Solid State Cavity QED System}
\author{Alexander Papageorge}
\email{papag@stanford.edu}
\affiliation{E. L. Ginzton Laboratory, Stanford University, Stanford, California 94305, USA}
\author{Arka Majumdar}
\affiliation{E. L. Ginzton Laboratory, Stanford University, Stanford, California 94305, USA}
\author{Erik D. Kim}
\affiliation{E. L. Ginzton Laboratory, Stanford University, Stanford, California 94305, USA}
\author{Jelena Vu\v{c}kovi\'c}
\affiliation{E. L. Ginzton Laboratory, Stanford University, Stanford, California 94305, USA}
\date{March 2011}
\begin{abstract}
The bichromatic driving of a solid state cavity quantum electrodynamics system is used to probe cavity
dressed state transitions and observe coherent interaction between the system and the light field. We theoretically demonstrate the higher order cavity-dressed states, supersplitting, and AC stark shift in a solid state system comprised of a quantum dot strongly coupled to a photonic crystal cavity for on- and far off-resonant cases. For the off-resonant case, phonons mediate off-resonant coupling between the quantum dot and the photonic resonator, a phenomenon unique to solid state cavity quantum electrodynamics.
\end{abstract}
\maketitle
\section{Introduction}

Many proposed methods in quantum information processing employ the strong optical nonlinearity created by a single quantum emitter coupled to an optical resonator \cite{kimble95}, as for example a quantum dot (QD) coupled to an optical microcavity \cite{vuckovic07}. Such solid state cavity quantum electrodynamics (QED) systems can be used for the scalable implementation of quantum information processing devices, but in order to do that it is important to observe the coherent interaction between the quantum dot and a laser field. The existence of the light-field dressed states manifests itself in several quantum optical phenomena. Examples include the Mollow triplet in resonance fluorescence measurements \cite{mollow69}, Rabi oscillations between the QD ground and excited states \cite{steel07}, and Autler-Townes splitting in the absorption spectrum of the QD \cite{AutlerTownes55}. 

The observation of the Mollow absorption spectrum is a classic example of bichromatic driving where a strong pump is used to dress the QD, whose absorption spectrum is observed via a weak probe \cite{mollow72}. Observing the Mollow triplet in the resonance fluorescence of the QD requires a very sensitive background-free measurement. Experiments performed in solid state systems have so far relied on complicated fabrication techniques to build structures where the probe and collected light follow orthogonal paths \cite{shih09}, or have used sophisticated signal isolation for background reduction \cite{atature09}.

Recently there has been a novel experimental demonstration that makes use of off-resonant QD-cavity coupling combined with bichromatic driving to observe the dressing of the QD \cite{vuckovic11.2}. When the QD is driven resonantly the cavity emits light through an incoherent read-out channel spectrally removed from the QD resonance \cite{finley08}, \cite{imamoglu07}, \cite{vuckovic10},  \cite{michler09}. This read-out channel arises from incoherent processes that result in the emission of photons at the cavity frequency under optical excitation of the QD and vice versa \cite{vuckovic11}, \cite{hughes11}.

In this paper we show that in addition to QD dressing, such bichromatic driving can be used to observe several coherent effects in a dressed cavity QED system, including higher order dressed states, supersplitting of the polariton states, and the AC stark shift of the dressed states for a resonant cavity-QD system. These observations are made by measuring changes in the cavity emission intensity as a function of the probe frequency.

\section{Theory}
\subsection{Physical Model}

We theoretically model driving the QD resonantly with a pump field strong enough to dress the exciton states. A weak probe beam is scanned across the QD resonance and the photoemission of the cavity is observed. We consider the optical transition of the QD as a two-level system, and model the coherent driving of a cavity QED system using the Jaynes-Cummings (JC) Hamiltonian
\begin{equation}\label{fishstew}
H=\omega_c a^\dag a + \omega_d \sigma^\dag \sigma + g(\sigma^\dag a + \sigma a^\dag) + J\Sigma + J^*\Sigma^\dag
\end{equation}
where $\omega_c$ and $\omega_d$ are, respectively, the cavity and QD resonant frequency, $a$ and $\sigma$ are, respectively, the annihilation operators for the cavity mode and the lowering operator for the QD, $g$ is half of the vacuum Rabi splitting, and $J$ is the Rabi frequency of the field driving the QD. $\Sigma$ is either $a$ or $\sigma$ depending on whether the cavity or QD is being driven, respectively. Since the laser field is bichromatic $J$ takes on the following form, where $\omega_l$ is the frequency of the pump laser, $J_1$ and $J_2$ are the Rabi frequencies of the pump and probe lasers respectively, and $\delta$ is the detuning between the pump and probe lasers
\begin{equation}
J=J_1 e^{i\omega_l t} + J_2 e^{i (\omega_l+\delta) t}
\end{equation}
Without loss of generality we assume that $J_1$ and $J_2$ are real. Pumping the QD on resonance ($\omega_l = \omega_d$) and transforming eq.~\eqref{fishstew} to a frame rotating with the pump laser field leads to
\begin{eqnarray}
\begin{split}
H&=\Delta_c a^\dag a +\Delta_d \sigma^\dagger\sigma+ g(\sigma^\dag a + \sigma a^\dag) + J_1\sigma_x +&\\  &J_2\left(e^{i\delta t}\sigma + e^{-i\delta t}\sigma^\dagger\right)&  \\ 
&\equiv H_0+ J_2 e^{i\delta t}\Sigma + J_2 e^{-i\delta t}\Sigma^\dagger&
\end{split}
\end{eqnarray}
where $\Delta_i=\omega_i-\omega_l$. The term $J_1\sigma_x + J_2\left(e^{i\delta t}\sigma + e^{-i\delta t}\sigma^\dagger\right)$ describes the bichromatic driving of the QD \cite{ficek93}, \cite{friedmann87}. While there is no frame in which the equation of motion is time-independent, the fact that the system is weakly probed allows us to solve the problem perturbatively by the method of continued fractions  \cite{ficek96}. From the Hamiltonian and associated incoherent loss terms we find the fluorescence spectrum of the cavity as a function of the pump strength $J_1$ and the pump-probe detuning $\delta$. To this end we develop a framework to calculate any number of observable quantities for a bichromatically driven cavity-dot system. 
To incorporate incoherent losses, the problem is framed in terms of the master equation for the density matrix
\begin{equation}\label{rhodot}
\begin{split}
\dot{\rho} = -i [H,\rho] + \mathcal{D}\left(\sqrt{2\gamma}\sigma\right)\rho + \mathcal{D}\left(\sqrt{2\kappa}a\right)\rho \\ + \mathcal{D}\left(\sqrt{2\gamma_d}\sigma^\dagger\sigma\right)\rho+ \mathcal{D}\left(\sqrt{2\gamma_r}a^\dagger\sigma\right)\rho
\end{split}
\end{equation}
where $\mathcal{D}\left(C\right)\rho$ indicates the Lindblad term $C\rho C^\dagger-\frac{1}{2}\left(C^\dagger C\rho + \rho C^\dagger C\right)$  associated with the collapse operator $C$. The second and third terms of eq.~\eqref{rhodot} represent cavity decay and spontaneous emission from the QD, with $\gamma$ and $\kappa$ being the spontaneous emission rate and cavity field decay rate, respectively. The fourth term induces pure dephasing, and represents a phenomenological interaction of the QD with its environment. The effect of pure dephasing is to broaden the resonant lineshapes and destroy coherence, decreasing the visibility of coherent effects \cite{lloyd09}. The term proportional to $\gamma_r$ is of particular importance as it describes phonon-mediated coupling between an off-resonant quantum dot and a cavity mode \cite{vuckovic11}, a phenomenon unique to solid-state systems where relaxation of the excited QD occurs through the generation or absorption of a phonon, and the creation of a photon in the cavity. Such a term better accounts for off-resonant coupling than pure dephasing alone as it induces population transfer between the QD and the cavity for larger detuning ranges, consistent with experiments \cite{vuckovic11}. As a simplification we take the low temperature limit of the phonon-mediated coupling, ignoring a term proportional to $\bar{n}a\sigma^\dagger$, where $\bar{n}$ is the population of phonons at frequency $\Delta=\Delta_c-\Delta_d$ as given by Bose-Einstein statistics. $\bar{n}$ is relevant in experimental systems as it is generally not negligible and gives rise to the temperature dependence of the off-resonant coupling. It should also be noted that here we only consider the case of a  QD blue-detuned from the cavity, where the relaxation of the QD corresponds to the creation of a phonon. The appropriate Lindblad terms would be different if the QD were red-detuned; specifically more terms would need to be included as off-resonant coupling is not observed in the zero temperature limit \cite{vuckovic11}. The master eq.~\eqref{rhodot} can be written in terms of Liouvillean superoperators as
\begin{equation}\label{rhoEOM}
\dot{\rho}=\left(\mathcal{L}_0+\mathcal{L}_+e^{i\delta t}+\mathcal{L}_-e^{-i \delta t}\right)\rho
\end{equation}
This formulation for the master equation is identical to eq.~\eqref{rhodot}. In the regime that we consider experimentally, $\mathcal{L}_\pm$ are proportional to $J_2$ and can be treated as perturbative additions to $\mathcal{L}_0$. Specifically,
\begin{equation}\label{therho0}
\begin{split}
\mathcal{L}_0\rho= -i [H_0,\rho] + \mathcal{D}\left(\sqrt{2\gamma}\sigma\right)\rho + \mathcal{D}\left(\sqrt{2\kappa}a\right)\rho \\ + \mathcal{D}\left(\sqrt{2\gamma_d}\sigma^\dagger\sigma\right)\rho + \mathcal{D}\left(\sqrt{2\gamma_r}a^\dagger\sigma\right)\rho
\end{split}
\end{equation}
\begin{equation}
\mathcal{L}_+\rho= -i J_2[\Sigma,\rho]
\end{equation}
\begin{equation}\label{therhominus}
\mathcal{L}_-\rho= -i J_2[\Sigma^\dagger,\rho]
\end{equation}
This equation can be solved by Floquet theory, and a solution of the form $\rho(t)=\displaystyle\sum_{n=-\infty}^{\infty}{\rho_n(t) e^{i n\delta t}}$ can immediately be postulated \cite{chiconeTEXT}. Introducing this trial solution to eq.~\eqref{rhoEOM}, taking the Laplace transform, and equating terms proportional to $e^{i n\delta t}$ yields the recurrence relation
\begin{multline}\label{MOCFeqn}
z\rho_n(z) +\rho(0)\updelta_{n0}+ in\delta\rho_n(t) \\ =  \mathcal{L}_0\rho_n(z) + \mathcal{L}_+\rho_{n-1}(z)+\mathcal{L}_-\rho_{n+1}(z)
\end{multline}
which can be solved numerically by the method of continued fractions. We seek the resonance fluorescence spectrum of the cavity which is found as the real part of the Fourier transform of the stationary two-time correlation function $\langle a^\dagger(t+\tau)a(t)\rangle$. Application of the quantum regression theorem allows this quantity to be calculated as $\operatorname{tr}\{a^\dagger M(\tau)\}$, where $M(\tau)$ solves the master equation with initial condition $M(0)=a\rho(t\rightarrow\infty)$ \cite{scullyzubairyTEXT}. From the recurrence relation and the aforementioned initial condition the method of continued fractions allows us to obtain an expansion of the Laplace transform of $M(\tau)$ of the form $M(z)=\displaystyle\sum_{n=-\infty}^{\infty} M_n(z+in\delta)$, from which the cavity resonance fluorescence spectrum is
\begin{equation}\label{bettyboop}
S(\omega)=\operatorname{Re}\left(\operatorname{tr}\{a^\dagger M_0(i\omega)\}\right)
\end{equation}
where $\omega$ is the angular frequency of the emitted light, centered at the frequency of the pump laser. In our calculation, $\rho_0$ is found to first order in $J_2$ by assuming all $\rho_n$ for $|n|>1$ are 0, reflecting the relatively weak probe strength. In the regime under consideration much less than one photon is ever in the cavity at any time (i.e. $\langle a^\dagger a \rangle \ll1$) and the photon basis is truncated to a small subspace of Fock states $\{|0\rangle, |1\rangle, |2\rangle\}$. These approximations are validated by observing no change in the calculation with an expansion of either basis.

\subsection{Method of Continued Fractions}

The method of continued fractions is performed by assuming the existence of matrices $\mathcal{S}_n$ and $\mathcal{T}_n$ with the following properties
\[
\rho_n=
  \begin{cases}
   \mathcal{S}_n\rho_{n-1} & \text{if } n > 0 \\
   \mathcal{T}_n\rho_{n+1} & \text{if } n < 0
  \end{cases}
\]
These matrices can be found explicitly by solving the following infinite system of equations.
\begin{equation}\label{defSN}
\mathcal{S}_n=-\left[\left(\mathcal{L}_0-\left(z+i n\delta\right)\mathds{1}\right) + \mathcal{L}_-S_{n+1}\right]^{-1}\mathcal{L}_+
\end{equation}
\begin{equation}\label{defTN}
\mathcal{T}_n=-\left[\left(\mathcal{L}_0-\left(z+in\delta\right)\mathds{1}\right) + \mathcal{L}_+T_{n-1}\right]^{-1}\mathcal{L}_-
\end{equation}
In practice, a solution is found by assuming that at some large $n$ ($-n$), the matrix $\mathcal{S}_n$ ($\mathcal{T}_{-n}$) is 0, and finding all other matrices. Once the set of $\mathcal{S}_n$ and $\mathcal{T}_n$ have been found, all $\rho_n$ can be found, starting with $\rho_0$
\begin{equation}\label{fishbrain}
\rho_0(z)=\left[\mathcal{L}_0-z\mathds{1} + \mathcal{L}_-\mathcal{S}_1 + \mathcal{L}_+\mathcal{T}_{-1}\right]^{-1}\rho(0)
\end{equation}
To obtain the resonance fluorescence spectrum, the initial conditions must be chosen carefully, and by setting $\rho(0)=a\rho(t\rightarrow\infty)$, then $M_0(z)=\rho_0(z)$ in eq.~\eqref{fishbrain}. The method of continued fractions is also used to calculate the steady state behavior of the density matrix, but in the long time limit, $\rho$ can be expanded as $\rho(t\rightarrow\infty)=\displaystyle\sum_{n=-\infty}^{\infty}{\rho_n e^{i n\delta t}}$, which is the same as the previous expansion but in this case the $\rho_n$ carry no explicit time dependence. The modified continued fractions matrices are found by setting $z=0$ in eqns. \eqref{defSN} and \eqref{defTN}. To lowest order, the steady state density matrix $\rho_{ss}=\rho_0$ is the nullspace of $\left(\mathcal{L}_0 + \mathcal{L}_-\mathcal{S}_1 + \mathcal{L}_+\mathcal{T}_{-1}\right)$. Normalizing $\rho_0$ such that its trace is 1, the density matrix to first order is $\left(\rho_0+\rho_1e^{i\delta t} +\rho_{-1}e^{-i\delta t}\right)/\left(1+\operatorname{tr}\{\rho_1\}e^{i\delta t} +\operatorname{tr}\{\rho_{-1}\}e^{-i\delta t}\right)$ which yields a first order time averaged density matrix $\rho_{ss} = \rho_0 - \operatorname{tr}\{\rho_{1}\}\rho_{-1} - \operatorname{tr}\{\rho_{-1}\}\rho_{1}$. This quantity is used to find the cavity resonance fluorescence spectrum with $\rho(0) = a\rho_{ss}$.

\section{Dressed State Probing and Supersplitting}

In the absence of dissipation or pure dephasing the Jaynes-Cummings Hamiltonian $H=\omega_c\left(a^\dagger a+\sigma^\dagger \sigma\right)+\Delta\sigma^\dagger\sigma+g\left(a^\dagger\sigma + a\sigma^\dagger\right)$ has an eigenvalue spectrum $n\omega_c+\frac{1}{2}\left(\Delta\pm\sqrt{4g^2n+\Delta^2}\right)$, where $\Delta=\omega_d-\omega_c$, and $n$ refers to the integer number of excitations in the system. A peak splitting can be observed in low power transmission, reflection, or absorption measurements, whose magnitude can be found by perturbation theory with a perturbing Hamiltonian $J\left(a e^{i\omega_l t} +h.c.\right)$. Under resonant excitation, the cavity transmission spectrum is 

\begin{equation}
\begin{split}
&T(\omega)\propto\\ &\frac{J^2\left(\gamma^2+(\Delta-\omega)^2\right)}{g^4+2g^2(\frac{1}{2}J^2 + \gamma\kappa+(\Delta-\omega)\omega)+(\gamma^2+(\Delta-\omega)^2)(J^2+\kappa^2+\omega^2)}
\end{split}
\end{equation}

which, for the case of zero detuning between the QD and cavity, has peaks at $\omega_{\pm}=\omega_c\pm\sqrt{\sqrt{g^2\left(g^2+J^2\right)+2g^2\gamma(\gamma+\kappa)}-\gamma^2}$. The dependence of the peak frequencies on the drive strength $J$ is the AC stark shift caused by coupling between the ground state and first manifold. As $J$ increases, higher orders in perturbation theory must be considered as the drive field couples higher order transitions between states in the JC manifold. As $J$ approaches the dressed state linewidth, these higher order transitions will become apparent in transmission spectra as additional resonances (higher order dressed states), and as $J$ surpasses the dressed state linewidth, dressing of the cavity dressed states will be observable as splitting of these polariton resonances (supersplitting). As we show below, all three effects, AC stark shift, higher order dressed states \cite{schoelkopf09}, and supersplitting \cite{rempe11}, \cite{bermanTEXT} can be observed by CW bichromatic driving of the cavity QED system.

In this simulation the cavity mode is resonantly pumped, and the transmission of a weak probe is used to provide an indication of the three aforementioned effects. The system is modeled by eqns.~\eqref{therho0}-\eqref{therhominus}, where $\Sigma=a$. The time-averaged cavity transmission is proportional to $\langle a^\dagger a\rangle = \operatorname{tr}\{a^\dagger a \rho_{ss}\}$, where $\rho_{ss}$.

\begin{figure}[htp]
\begin{center}
\includegraphics[scale=.25]{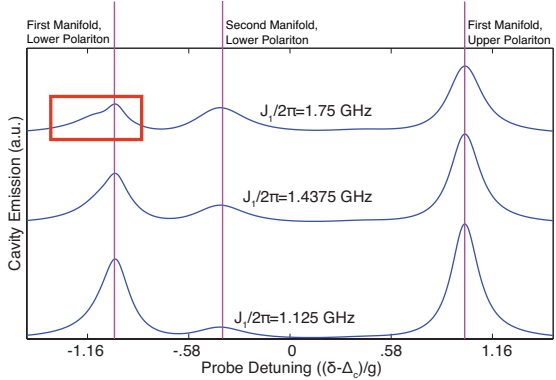}
\end{center}
\caption{Deviation of cavity emission from steady state for increasing values of the pump Rabi frequency $J_1$ under bichromatic driving wherein the pump is resonant to the lower polariton of the first manifold ($\omega_l=\omega_-$). Plots are vertically offset for clarity. Parameter values used in the simulation are $\gamma/2\pi=1$ GHz, $\gamma_d/2\pi=1$ GHz, $\gamma_r/2\pi=0$, $\kappa/2\pi=3$ GHz, $\omega_c=\omega_d$, $g/2\pi=30$ GHz, $J_2/2\pi=.01$ GHz. The box identifies the onset of supersplitting for the upper polariton.}
\label{PinkDresses}
\end{figure}

We assume that a cavity with Q$\approx$54000 is resonant with a QD ($\omega_c=\omega_d$), and the pump laser is tuned to $\omega_l=\omega_-$, maximizing the field inside the cavity. This corresponds to $\kappa/2\pi=3$ GHz, $\gamma/2\pi=1$ GHz, $\gamma_d/2\pi=1$ GHz, and $g/2\pi=30$ GHz. These parameters represent optimistic but realizable photonic crystal cavities made of GaAs containing InAs QDs. A weak probe is swept across the cavity/QD resonance and the total emission intensity of the cavity (proportional to $\langle a^\dagger a\rangle$) is measured. Fig. \ref{PinkDresses} displays the cavity transmission spectrum as a function of the probe wavelength for increasing pump power $J_1$. The probe is weak and equal to $J_2/2\pi=.01$. The pump is resonant with the lower polariton of the first manifold, at a frequency $\omega_-$. The vacuum Rabi splitting is clearly demonstrated in peaks located at $\sim\omega_c\pm g$, while the third visible peak is located at $\sim\omega_c-(\sqrt{2}-1)g$, and is indicative of a transition between the first and second manifolds, as the combined lasers make up an energy  $\sim2\omega_c+\sqrt{2}g$. Increasing the pump Rabi frequency past a threshold value induces dressing of the polariton states, visible in the red box in the uppermost curve of Fig. \ref{PinkDresses} as splitting in the transmission spectrum. Pure dephasing effectively increases the value of $J_1$ necessary for supersplitting to be observed.

\begin{figure}[htp]
\begin{center}
\includegraphics[scale=.25]{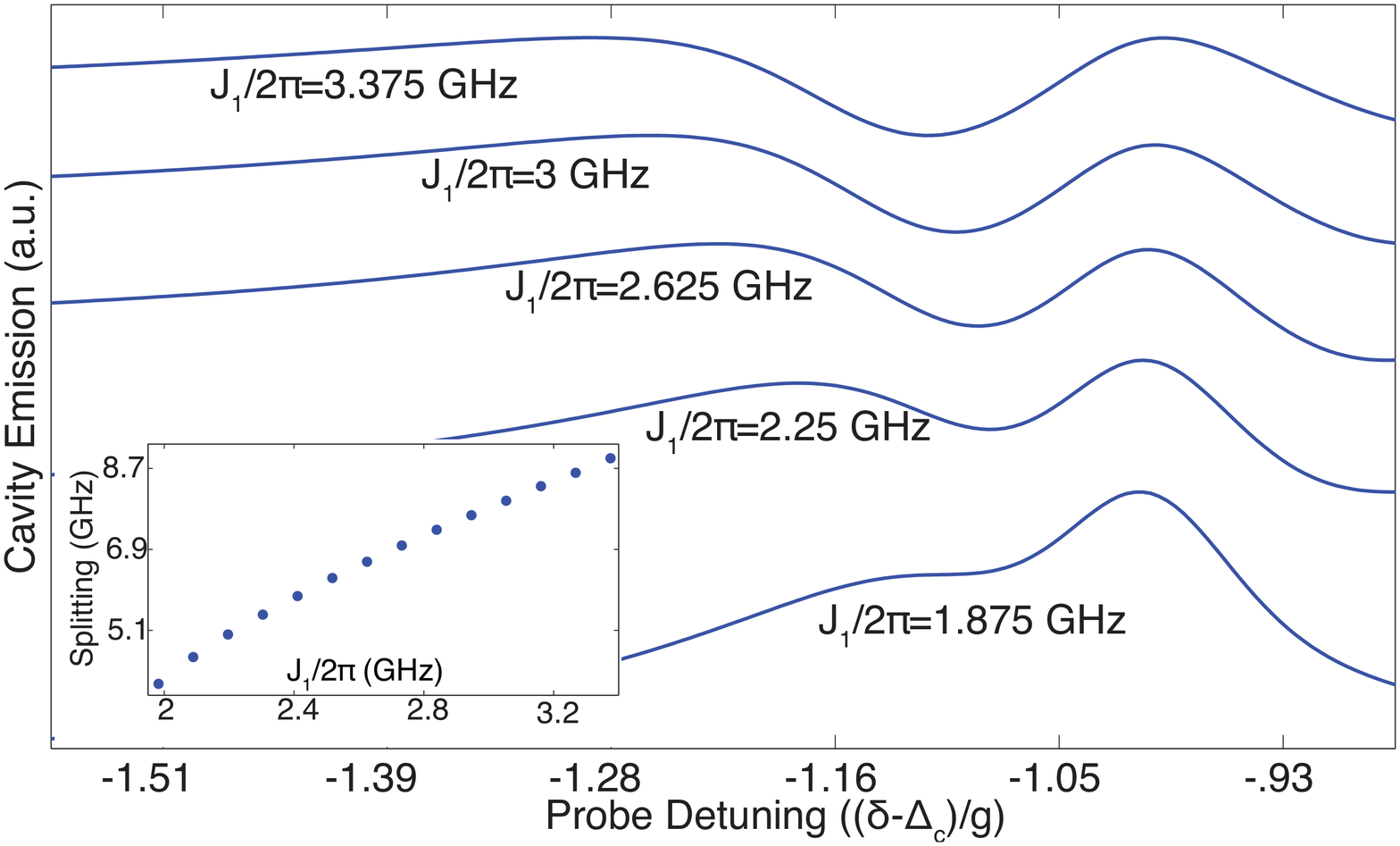}
\end{center}
\caption{Increased supersplitting (red box in Fig. \ref{PinkDresses}) for increasing values of the pump Rabi frequency $J_1$. Parameter values used in the simulation are the same as those used in Fig. \ref{PinkDresses}.}
\label{OrangeDresses}
\end{figure}

Supersplitting in the transmission spectrum occurs when the field radiated by the polariton destructively interferes with the pump field. It can at first be described clasically and occurs even when $J_1$ is below the polariton linewidths, which are approximately 2 GHz in our simulations. Fig. \ref{OrangeDresses} shows the increasing splitting of the driven polariton with pump power. The splitting is expected to be linear in $J_1$ for a two-level system, but the influence of the higher order states complicates the situation and alters the functional dependence.

\begin{figure}[htp]
\begin{center}
\includegraphics[scale=.25]{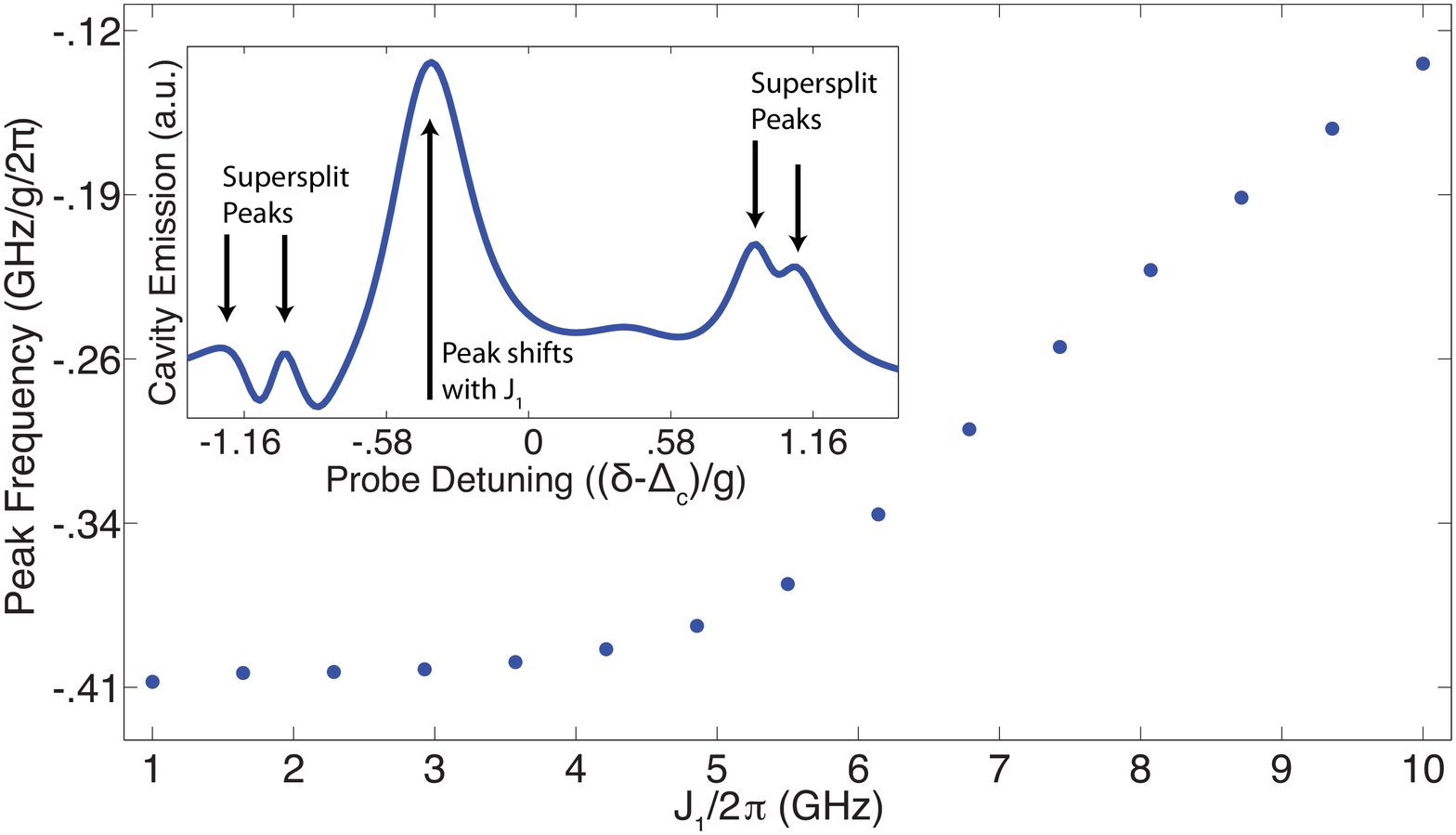}
\end{center}
\caption{Wavelength (relative to QD resonance) of the emission peak from the second manifold polariton (see inset) for increasing values of the pump Rabi frequency $J_1$. Parameter values used in the simulation are the same as those used in Fig. \ref{PinkDresses}. Inset shows the cavity emission spectrum for $J_1/2\pi=3$ GHz. Supersplitting is visible in both first manifold polaritons, and the two photon transition to the second manifold is large.}
\label{GreenDresses}
\end{figure}

At higher pump powers, when both first manifold polaritons display splitting, the second manifold states exhibit a notable AC stark shift. Fig. \ref{GreenDresses} shows the cavity emission spectrum for a value of $J_1$ large enough to split both first manifold polaritons. Both polaritons in the second manifold are visible, one of which is significant. When the pump and probe frequencies satisfy the two-photon resonance condition for the second manifold, and the pump Rabi frequency surpasses the loss rate of the second manifold, the polaritons in the second manifold become increasingly visible in the transmission spectrum. Increasing the pump Rabi frequency, the lower polariton in the second manifold displays a notable AC stark shift as seen in Fig. \ref{GreenDresses}. The resonance shift displays a clear transition when $J_1/2\pi\approx5$ GHz, when the pump Rabi frequency surpasses the loss rate of the second manifold.

The previously described simulations place the QD and cavity resonances at the same frequency; if they were detuned the coherent effects would be less visible, but all plots would appear qualitatively the same.

\section{Observation of QD Dressed States}

\begin{figure}[htp]
\begin{center}
\includegraphics[scale=.3]{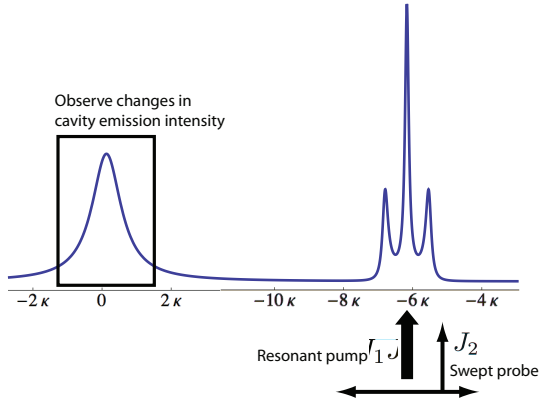}
\end{center}
\caption{A schematic of the experimental setup for bichromatic driving of the off-resonant QD-cavity system. The emission of the cavity is observed. The QD is pumped resonantly with a strong laser. A weak probe is swept across the resonance while the intensity of the cavity emission is measured in response.}
\label{falagfel}
\end{figure}

In the experiment we emulate \cite{vuckovic11.2} (see Fig. \ref{falagfel}), a QD coupled to an off-resonant photonic crystal cavity is pumped resonantly. The light emitted from the system is dispersed by a grating and the signal at the cavity frequency is spectrally isolated. A weak probe beam is scanned across the QD resonance and the change in cavity emission intensity is measured. Spectra obtained in this manner change dramatically as the pump power is increased and the QD states are dressed by the pump laser. We calculate the cavity response as the maximum value of the cavity emission spectrum at a frequency nearest to the native cavity frequency, mimicking the experimental measurement. Our simulations were performed with experimentally relevant parameters for InAs QDs coupled to GaAs photonic crystal cavities \cite{vuckovic10.2}. We ignore coherent cavity-QD coupling and set $g=0$. Not only does this make the underlying physics easier to understand, but off-resonant coupling is usually observed in weakly coupled systems. As will be discussed later, the observable effect of $g$ is to create an asymmetry in the cavity intensity lineshape. The other parameters are representative of experiments in our group, with $\gamma/2\pi=1$ GHz, $\gamma_d/2\pi=3$ GHz, $\kappa/2\pi=17$ GHz, $\Delta=\omega_d-\omega_c=8\kappa$. The strength of the off resonant coupling has been approximated as $\gamma_r/2\pi=.1$ GHz to qualitatively match emission spectra observed in experiment.

Performing the simulation for the full system involves finding the cavity resonance fluorescence spectrum from eq.~\eqref{rhodot} and observing how the peak of the cavity emission changes as a function of the pump-probe detuning $\delta$. The physics can be understood intuitively by considering the experiment as probing a four level system created by the resonant pump beam dressing the QD. When the probe beam is on resonance with one of the dressed state transitions, coherent effects alter the cavity resonance fluorescence spectrum significantly as the pump field $J_1$ approaches and surpasses the natural linewidth of the QD.

The effect of the collapse operator $a^\dagger\sigma$ is to move population from the QD to the cavity. Thus, changes in cavity emission are caused by changes in the excited state population of the QD as induced by the probe. Because the probe has the effect of moving population, the measurement is similar to an absorption measurement. Interference between quantum mechanical pathways counters the effect of the probe moving population into the QD excited state, resulting in characteristic dips in the cavity emission intensity lineshape. In essence, the probe field has the effect of altering the steady state QD excited state population, and it is the interference of quantum mechanical pathways that results in the observed lineshapes, similar to an absorption measurement \cite{mollow72}.

\begin{figure}[htp]
\begin{center}
\includegraphics[scale=.15]{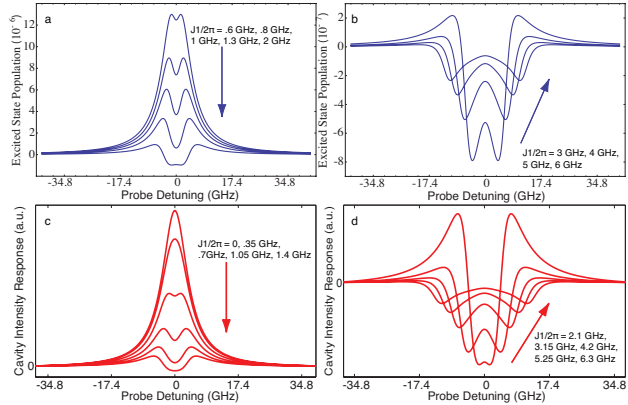}
\end{center}
\caption{(a)-(b) Excited state population of a probed two level system under bichromatic driving (see Fig. 4). The values of the parameters are: $\gamma/2\pi$=1 GHz, $J_2/\gamma=.01$ GHz, $\gamma_d/\gamma=3$ GHz. (c)-(d) Deviation of off-resonant cavity emission from steady state for different values of pump powers. Parameter values used in the simulation are $\gamma/2\pi=1$ GHz, $\gamma_d/2\pi=3$ GHz, $\gamma_r/2\pi=.1$ GHz, $g/2\pi=0$, $\kappa/2\pi=17$ GHz, $\Delta=\omega_d-\omega_c=8\kappa$, $J_2/2\pi=.35$ GHz.}
\label{mackerel}
\end{figure}

For a Rabi frequency lower than the QD spontaneous emission rate, $2J_1\textless \gamma$, the QD dressed states are not discernible, and the cavity emission lineshape (i.e., cavity emission intensity vs pump-probe detuning $\delta$) is a simple Lorentzian. The linewidth of this Lorentzian is approximately the natural QD linewidth adjusted by power broadening and pure dephasing. As the Rabi frequency increases beyond a critical threshold $2J_1\sim\gamma$, a notable change in the cavity emission response occurs. Two dips appear symmetrically around $\delta=0$ that deepen and separate further as the pump power is increased. These dips are direct evidence of the dressing of the QD, and are separated by twice the Rabi frequency. The response lineshape is directly related to the excited state population of the dot, given to second order in the probe strength as
\begin{widetext}
\begin{align}
\rho_{ee}=&\frac{J_1^2}{2J_1^2+ \gamma\left(\gamma+\gamma_d\right)}+\nonumber \gamma J_2^2\times \\  &\frac{\left(8J_1^4\left(\gamma+\gamma_d\right)\left(-2\left(\gamma+\gamma_d\right)^2-3\delta^2\right) + \left(\gamma+\gamma_d\right)^3\left(4\gamma^2+\delta^2\right)\left(\left(\gamma+\gamma_d\right)^2+\delta^2\right) + 4 J_1^2\delta^2\left(-3\gamma\left(\gamma+\gamma_d\right)^2 + \gamma_d\delta^2\right)\right)}{\left(2J_1^2+\gamma\left(\gamma+\gamma_d\right)\right)^2\left(\left(\gamma+\gamma_d\right)^2+\delta^2\right)\left(4\left(2J_1^2+\gamma\left(\gamma+\gamma_d\right)\right)^2+\left(-8J_1^2+5\gamma^2+2\gamma\gamma_d+\gamma_d^2\right)\delta^2+\delta^4\right)}
\end{align}
\end{widetext}
Fig. \ref{mackerel}a-b shows the steady state excited state population of a QD under bichromatic driving. The population is given with respect to the unprobed value, $J_1^2/(2 J_1^2 + \gamma^2 + \gamma \gamma_d)$. For a weak pump, the response of the QD excited state is Lorentzian; for a strong pump, $2J_1\sim\gamma$, the QD begins to saturate at which point the dressed states can be resolved. When the probe is resonant with a dressed state transition the excited state population is decreased from its steady state (probe absent) value. This effect gives rise to the central dip observed in Fig. \ref{mackerel}a-b. For increasing $J_1$ the dip separates into two distinct dips, each of which corresponds to one of the two dressed state transitions. In the absence of pure dephasing, the individual dips can be resolved for smaller values of $J_1$, and the threshold value of $J_1$ also decreases.

Fig. \ref{mackerel}c-d shows the change in off-resonant cavity emission from the steady state value as a function of the pump-probe detuning for various values of pump Rabi frequency $J_1$. The lineshapes reflect the excited state of the QD, broadened slightly by the cavity linewidth. The magnitude of the cavity response decreases with increasing pump power since the $J_1$ dependent background is subtracted from each curve. The background increases with pump power, saturating with the QD excited state population. Thus, the probe makes a decreasing contribution to the total incident power, effectively moving less population into the excited state and producing a smaller overall effect. In the perturbative limit we are considering, increasing the probe power simply increases the overall visibility of the signal as the cavity response is proportional to $J_2$.

It should be stressed that in the absence of coherent coupling the intensity of the cavity emission is dependent only on $\gamma_r$, which describes the strength of the phonon assisted process, and not explicitly on the detuning. Since the incoherent coupling is phonon-mediated, the coupling constant should depend on the population of phonons at the energy of the detuning, ie $\gamma_r\propto\bar{n}_b=(\operatorname{exp}(\hbar|\Delta|\beta)-1)^{-1}$. Our model predicts that the cavity emission intensity should be roughly proportional to $\gamma_r(T,\Delta)$. The factor of proportionality, as derived in \cite{vuckovic11}, is a complicated summation over virtual state coupling strengths. As a point of reference, at 10K, $\bar{n}_b$ for phonons of frequency 136 GHz is approximately 1.

\begin{figure}[htp]
\begin{center}
\includegraphics[scale=.25]{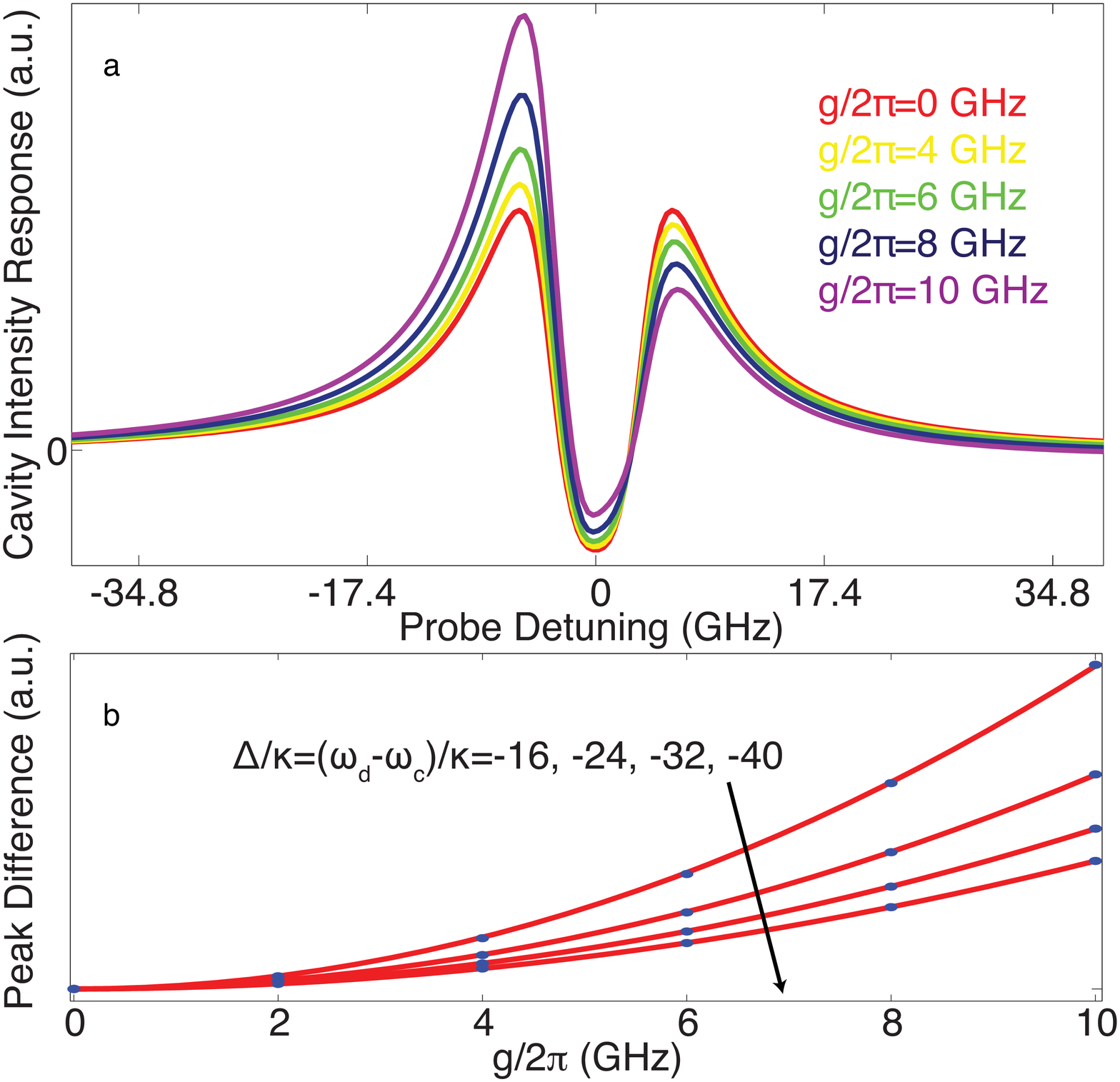}
\end{center}
\caption{(a) Deviation of cavity intensity from steady state for different values of the QD-cavity splitting $g$. Parameter values used in the simulation are $\gamma/2\pi=1$ GHz, $\gamma_d/2\pi=3$ GHz, $\gamma_r/2\pi=.1$ GHz, $\kappa/2\pi=17$ GHz, $\Delta=\omega_d-\omega_c=8\kappa$, $J_2/2\pi=.35$ GHz, $J_1/2\pi=1.75$ GHz. (b) Dependence of the difference in intensity between the two peaks on $g$ and $\Delta$. The remaining parameters are the same as in (a). Dots show simulation results, curves show fits to $g^2/(\alpha +\Delta)$ for free parameter $\alpha$. All values are in GHz.}
\label{casket}
\end{figure}

The effect of the coherent coupling $g$, which was excluded in the previous calculations for simplicity, is to create an asymmetry between the features of the cavity response lineshape. Far off resonance, the cavity coupling enhances the QD resonance fluorescence, and this effect is stronger for the dressed state transition nearer in frequency to the cavity resonance. The cavity response when this dressed state is probed is suppressed relative to the background, while the response to probing the other dressed state is enhanced. This asymmetry is shown in Fig. \ref{casket}a where the cavity response is shown for increasing $g$. In this regime where $\Delta\gg\kappa,\gamma$ the difference in the peak intensities is proportional to $J_2 \gamma_r g^2/(\alpha + \Delta)$, for an empirical fitting parameter $\alpha$. Fig. \ref{casket}b shows this functional dependence.

By tuning the temperature of the QD-cavity system, the relevant parameters can be found experimentally and $g$ can be determined. One difficulty is that $\gamma_r$ is also dependent on temperature and detuning. An alternative, but equivalent method is to fit the ratio of the larger peak intensity to the smaller peak intensity to a function of the form $2\alpha x/(1+\beta - \alpha x)$, where $\alpha$, $\beta$ are fitting parameters, and $x=g^2/\Delta$. In making this measurement it should be noted that asymmetries in the lineshape can also be caused by driving the QD with a detuned laser, and thus the pump laser should be very carefully tuned to the QD resonance.

\section{Conclusion}

In this paper we have theoretically analyzed the observation of dressed and redressed states in a solid state cavity QED system by a bichromatic CW pump-probe experiment. We have shown that the higher order polaritons will be visible in such transmission measurements for the current system parameters. By increasing the pump power, AC stark shift and supersplitting of the polariton resonances can be observed, an indication of dressing of the polariton states. Additionally, bichromatic driving of the QD can be used to observe the dressing of the QD through an incoherent off-resonant QD-cavity coupling unique to solid state systems. Using the off-resonant cavity to make spectroscopic measurements of the QD could enable a more convenient method of reading the state of the quantum dot in quantum information processing applications.

The authors acknowledge financial support provided by the the National Science Foundation (NSF Grant No. DMR-0757112), the Army Research Office (Grant No. W911NF-08-1-0399) and the Office of Naval Research (PECASE Award, N00014-08-1-0561). A.M. was supported by the Stanford Graduate Fellowship (Texas Instruments fellowship). E.K. was supported by the Intelligence Community (IC) Postdoctoral Research Fellowship.

\bibliographystyle{unsrt}
\bibliography{BichromaticQDref}
\end{document}